\documentclass[intlimits,twoside,a4paper]{article}

\usepackage[cp1251]{inputenc}

\usepackage[eqsecnum]{cmpj3}

\usepackage{graphics}

\usepackage{bm}
\issue{2020}{23}{2}{23501}
\doinumber{10.5488/CMP.23.23501}

\title[Electromagnetic field energy in an absorptive medium with
	temporal and spatial dispersion]%
{Electromagnetic field energy in an absorptive medium with
	temporal and spatial dispersion
}
\author[A.G.~Zagorodny,  S.A.~Trigger, A.I.~Momot]{A.G. Zagorodny\refaddr{label1},
        S.A. Trigger\refaddr{label2}, A.I.~Momot\refaddr{label3}}
\addresses{
\addr{label1} Bogolyubov Institute for Theoretical Physics of the National Academy
	of Sciences of Ukraine, \\ 	14-b Metrolohichna St., 03680 Kyiv, Ukraine
\addr{label2} Joint Institute for High Temperatures, Russian Academy of Sciences,\\
	13 Izhorskaya St., Bd. 2, 125412 Moscow, Russia
\addr{label3} Taras Shevchenko National University of Kyiv, 
              64/13 Volodymyrs'ka St., 01601 Kyiv, Ukraine
}

\def\J{\mathbf{J}}
\def\r{\mathbf{r}}
\def\v{\mathbf{v}}
\def\k{\mathbf{k}}

\def\u{\mathbf{u}}
\def\P{\mathbf{P}}
\def\E{\mathbf{E}}
\def\B{\mathbf{B}}
\def\H{\mathbf{H}}
\def\D{\mathbf{D}}
\def\R{\mathbf{R}}

\def\F{\mathbf{F}}

\def\ext{\mathrm{ext}}
\def\max{\mathrm{max}}
\def\EK{\mathrm{EK}}
\def\eff{\mathrm{eff}}
\def\e{\mathrm{e}}

\def\reg{\mathrm{reg}}
\def\I{\mathrm{I}}
\def\L{\mathrm{L}}
\def\T{\mathrm{T}}

\DeclareMathOperator{\rot}{rot}
\DeclareMathOperator{\rdiv}{div}

\newcommand{\cv}{v}

\date{Received February 27, 2020}

\begin{document}

\maketitle

\begin{abstract}
General relations for electromagnetic field energy outside the transparency domain are proposed. It is shown that charged particle contribution to the energy of electromagnetic perturbations in the general case can be described in terms of a bilinear combination of the dielectric polarizability of the medium. The explicit form of such contribution is found. The relations obtained are used to generalize the Planck law to the case of an absorptive medium.


\end{abstract}

\section{Introduction}
The energy density of an electromagnetic wave in a medium with spatial and temporal dispersion can be consistently defined only in the transparency domain \cite{1,2,3,4,5}. After the pioneer Brillouin result for the electromagnetic wave energy in dispersive transparent media \cite{6,7} a lot of papers have been published on this subject (see, for example, \cite{8,9,10,11,12,13,14,15,16,17,18,19,20,21}) and many attempts to generalize the Brillouin approach have been made \cite{10,16,18,19,20,21} to take into account absorptive properties of the medium. Nevertheless, the results known from the literature do not concern the general solution of the problem, but only various particular cases (weak absorption \cite{10,16,18,20,21}, medium with no spatial dispersion \cite{11,15,16,17,18,19,20,21}, some specific field configuration, particularly the two-wave construction to extend the Brillouin approach \cite{19} etc.). In contrast to these results, in the present paper we propose general relations to describe the energy density without such restrictions for any field configuration.

As is known, the energy of an electromagnetic perturbation in a matter contains the ``pure'' electromagnetic energy and the kinetic energy of charge carriers obtained due to their motion in the electromagnetic field \cite{2,3,10}. If neutral particles (i.e., atoms or molecules) are present, the additional potential energy acquired by bound electrons in such a field  should be also added
\cite{2,10,11,12,14,15,17,18}. Beside that, in the case of absorptive medium some part of electromagnetic energy is converted into  heat~\cite{10,12,14,18}. Thus, the problem arises to consistently  describe all these quantities. This introduces principal difficulties into generalizing the Brillouin formula to the case of dispersive absorptive medium since in such a case the macroscopic Maxwell equations generate a Poynting-like equation that does not provide an explicit identification of the total energy of electromagnetic perturbations and especially the heat production in contrast to the case of an nondispersive medium for which the total energy of the field is well defined and the heat production is absent.

In order to avoid the above-mentioned difficulties, it is possible to calculate all constituents of the electromagnetic field energy directly and express them in terms of dielectric susceptibilities as it was done for the case of a dissipative medium without spatial dispersion  \cite{2,10,11,14,15,17,18}. This approach can be justified using the energy balance equation which follows from the combination of the Maxwell equations and the kinetic equation for charge carriers. Such energy balance equation  was originally formulated by Ginzburg for a plasma medium \cite{8,9}. Similar energy balance equation can be formulated for the combined plasma-molecular medium using the kinetic equations for free and bound charged particles \cite{22,23,24}. In spite of the fact that the general ideas of electromagnetic field energy description were formulated many years ago, it was not yet applied to the case of absorptive medium with spatial dispersion.

The purpose of the present contribution is to derive a general relation for the energy of electromagnetic perturbation in the medium with temporal and spatial dispersion. We use the idea proposed in \cite{8,9,11}, namely, we treat the energy of the perturbation as a sum of the electromagnetic field energy and particle energy (both kinetic and potential) acquired by the particles in the field. The relations thus obtained are applied to calculate the fluctuation field energy and to generalize the Planck formula for the case of non-transparent medium with spatial and temporal dispersion.

The paper is organized in the following manner. In section~\ref{sec2} we formulate the basic equations and discuss the general statement of the problem. Section~\ref{sec3} gives the derivation of the kinetic and potential particle energy in the presence of an electromagnetic perturbation. We show that the results obtained make it possible to recover the relations known from the literature for various particular cases. In section~\ref{sec4} we apply the obtained formulae to the description of the fluctuation field energy density in the case of non-transparent medium with spatial and temporal dispersion. The results of numerical calculations concerning the effect of a nontransparent medium on the energy spectrum are presented in section~\ref{sec5}.

\section{Basic set of equations and statement of the problem} \label{sec2}

We start from the Maxwell equations for the electromagnetic field in a medium
in the form that is often used in the plasma theory \cite{3,4,25,26}
\begin{eqnarray}
\rot \E (\r,t) &=& - \frac{1}{c} \frac{\partial \B(\r,t)}{\partial t}\,, \nonumber\\
\rdiv \B(\r,t) &=& 0, \nonumber\\
\rot \B(\r,t) &=& \frac{1}{c} \frac{\partial \D(\r,t)}{\partial t} +
\frac{4\piup}{c} \J^\text{e}(\r,t), \nonumber\\
\rdiv \D(\r,t) &=& 4\piup \rho^\text{e}(\r,t),
\label{eq2-1}
\end{eqnarray}
where $\J^\text{e}(\r,t)$ and $\rho^\text{e}(\r,t)$ are  the external sources, if present. In the case under consideration $\H(\r,t)\equiv\B(\r,t)$,  and thus the total medium response to the electromagnetic field is described by the dielectric permittivity tensor $\varepsilon_{ij}(\r,\r';t-t')$
\begin{equation*}
	D_i(\r,t) = E_i(\r,t)+4\piup\int\limits_{-\infty}^t \rd t' J_i(\r,t')
	= \int\limits_{-\infty}^t \rd t' \int \rd\r' \varepsilon_{ij} (\r,\r';t-t') E_j(\r',t')\,,
\end{equation*}
where $J_i(\r,t)$ is the total induced current that includes all kinds of responses and can be expressed in terms of the conductivity tensor  $\sigma_{ij}(\r,\r';t-t')$ \cite{26}
\begin{eqnarray}
J_i(\r,t)=\int\limits_{-\infty}^t \rd t' \int \rd\r' \sigma_{ij}(\r,\r';t-t')E_j(\r',t').
\label{eq2-2}
\end{eqnarray}
Thus,
\begin{eqnarray}
\varepsilon_{ij}(\r,\r';t-t')=\delta_{ij} \delta(\r-\r') \delta(t-t')
+ 4\piup\int\limits_{t'}^t \rd t'' \sigma_{ij}(\r,\r';t''-t').
\label{eq2-3}
\end{eqnarray}

In the case of spatially homogeneous stationary medium with temporal and spatial dispersion $\varepsilon_{ij}(\r,\r';t-t') =\varepsilon_{ij}(\r-\r';t-t')$ and
\begin{eqnarray}
D_{i\k\omega} & =&\varepsilon_{ij}(\k,\omega)E_{j\k\omega}\,, \nonumber\\
\varepsilon_{ij}(\k,\omega) & =&\int\limits_0^\infty \rd\tau \,\re^{\ri\omega\tau} \int \rd\R\,\re^{-\ri\k\R} \varepsilon_{ij}(\R,\tau), \qquad \R = \r-\r', \quad \tau=t-t'.
\label{eq2-4}
\end{eqnarray}
In this case, the dispersion relation for eigenfrequency of a free electromagnetic wave is given by
\begin{equation}\label{eq2-5}
\det\Lambda_{ij}(\k,\omega(\k))=0,
\end{equation}
where
\[\Lambda_{ij}(\k,\omega)=\varepsilon_{ij}(\k,\omega)-\frac{k^2c^2}{\omega^2} \left(\delta_{ij}-\frac{k_ik_j}{k^2}\right).
\]

We also need  equations describing the interaction of electromagnetic fields with the medium. In what follows we illustrate the possibility to calculate the energy of electromagnetic perturbation using a plasma-like medium. Thus, we supplement equations (\ref{eq2-1}), (\ref{eq2-2}) with the kinetic equation for plasma particles
\begin{eqnarray}
\left\{ \frac{\partial}{\partial t} +\v\frac{\partial}{\partial \r} + \frac{e_\alpha}{m_\alpha} \F^{\ext} + \frac{e_\alpha}{m_\alpha} \left[\E(\r,t) +\frac{\v}{c} \times \B(\r,t)\right] \frac{\partial}{\partial \v} \right\}\ f_\alpha(\r,\v,t) = \I_\alpha\,,
\label{eq2-6}
\end{eqnarray}
where $f_\alpha(\r,\v,t)$ is the distribution function of particles of $\alpha$ species, $\I_\alpha$ is the collision term, $\F^{\ext}$  is the external force field, if present, the other notations are traditional.

Equation (\ref{eq2-6}) is valid in the case of a classical plasma-like medium. Appropriate calculations for the case of a combined plasma-molecular medium can be performed using the model of bound particles (see, for instance, \cite{10,11,12,14,15}). Quantum description of both plasma and plasma-molecular systems is also possible \cite{22,23,24, 27,28}. However, since the formulation of the general approach does not require the explicit form of the response function (except for the calculation of specific examples) as is shown below, we need to know only the general relation between the induced macroscopic currents $\J(\r,t)$ and the self-consistent electric field $\E(\r,t)$ given by equation (\ref{eq2-2}).

Using equations (\ref{eq2-1}) we obtain the well-known equation
\begin{equation}
\frac{1}{4\piup} \left(\E\frac{\partial \D}{\partial t} +\B\frac{\partial\B}{\partial t}\right) +\J^\text{e} \E=-\frac{c}{4\piup} \rdiv[\E\B],
\label{eq2-7}
\end{equation}
which is reduced to the Poynting equation in the case of non-dispersive ($\D=\varepsilon\E$) medium
\begin{equation}
\frac{\partial}{\partial t}\left( \frac{\varepsilon\E^2+\B^2}{8\piup}\right) +\J^\text{e}\E =-\frac{c}{4\piup} \rdiv[\E\B].
\label{eq2-8}
\end{equation}
According to the traditional interpretation, the term in the brackets is nothing else but the field energy density consisting of the electric and magnetic parts
\[
W=W_E+W_B\,,
\]
where
\[
W_E=\frac{1}{8\piup}\, \varepsilon\E^2=\frac{1}{8\piup}\E\D, \qquad W_B=\frac{1}{8\piup}\B^2.
\]
Taking into account that $\D=\E+4\piup\P$, where $\P$ is the polarization vector, it is easy to see that the electric field energy includes the energy of ``pure'' electric field and the energy of dipoles in the self-consistent electric field
\[
W_E=\frac{1}{8\piup}(\E^2+4\piup\E\P).
\]

In the case of quasi-monochromatic field, it is possible to show that \cite{10}
\begin{eqnarray}
\frac{1}{4\piup}\frac{\partial\D}{\partial t}\E &=& \frac{\partial W_E}{\partial t}+Q = \frac{1}{16\piup} \frac{\partial}{\partial\omega}\left[\omega \varepsilon'_{ij}(\k,\omega)\right]\frac{\partial}{\partial t}
E_{\k i}E_{\k j}^* + \frac{\omega \varepsilon''_{ij}(\k,\omega)}{8\piup} E_{0i} E_{0j}^* \nonumber\\
&+& \frac{1}{16\piup} \frac{\partial \varepsilon''_{ij}(\k,\omega)}{\partial\omega} \left(\frac{\partial E_{0i}}{\partial t} E_{0j}^* -\frac{\partial E_{0j}^*}{\partial t} E_{0i}\right) \nonumber\\
&-& \frac{\ri\omega}{16\piup} \frac{\partial\varepsilon''_{ij}(\k,\omega)}{\partial k_l}
\left(\frac{\partial E_{0i}}{\partial r_l}E_{0j}^* -\frac{\partial E_{0i}^*}{\partial r_l} E_{0j}\right),
\label{eq2-9}
\end{eqnarray}
where $\varepsilon'_{ij}(\k,\omega)$, $\varepsilon''_{ij}(\k,\omega)$ are the real and imaginary parts of the dielectric permittivity tensor. Here, $\E_0$ is assumed to be weakly dependent on time and coordinate.

In the transparency domain, one obtains the well-known Brillouin formula
\begin{equation}
W=\frac{1}{16\piup} \left\{ \frac{\partial}{\partial\omega} [\omega \varepsilon_{ij}(\k,\omega)]E_{\k i}E_{\k j}^*+B_{\k i} B_{\k j}^*\right\}.
\label{eq2-10}
\end{equation}
However, equations (\ref{eq2-9}), (\ref{eq2-10}) cannot be used in the case of a strongly absorptive medium. To get rid  of this restriction we can use the idea to derive an equation for energy balance which  explicitly takes into account the particle energy. This idea was suggested by Ginzburg \cite{8,9}. In order to derive such an equation it is necessary to multiply the kinetic equation (\ref{eq2-6}) by $n_\alpha m_\alpha \cv^2/2$ and integrate over the velocity $\v$. The result is
\begin{align}
\frac{\partial}{\partial t}\int \rd\v \frac{n_\alpha m_\alpha \cv^2}{2} f_\alpha(X,t) +\frac{\partial}{\partial\r} \int \rd\v\, &\v \frac{n_\alpha m_\alpha \cv^2}{2} f_\alpha(X,t) 
+ \int \rd\v \frac{n_\alpha e_\alpha \cv^2}{2} \left[\E+\frac{\v}{c}\times\B\right] \frac{\partial f_\alpha(X,t)}{\partial\v}=0;\\
& \int \rd\v\,\I_\alpha=0.\nonumber
\label{eq2-11}
\end{align}
Taking into account that
\begin{eqnarray}
\sum_\alpha \int \rd\v \frac{n_\alpha e_\alpha \cv^2}{2}\left[\E+\frac{\v}{c}\times\B\right] \frac{\partial f_\alpha(X,t)}{\partial\v}= -e_\alpha n_\alpha \int \rd\v\, \v\E f_\alpha(X,t)= -\E\J
\label{eq2-12}
\end{eqnarray}
and combining equation (\ref{eq2-12}) with the equation~(\ref{eq2-7}), which can be written in the form
\begin{eqnarray}
\frac{1}{4\piup} \left(\E\frac{\partial\E}{\partial t}+\B\frac{\partial\B}{\partial t}\right) + \E\J +\J^\text{e}\E=-\frac{c}{4\piup}\rdiv[\E\B],
\label{eq2-13}
\end{eqnarray}
one obtains the equation for the energy balance
\begin{eqnarray}
&&\frac{\partial}{\partial t}\left\{ \frac{1}{8\piup} \left[\E^2(\r,t)+\B^2(\r,t)\right]+
\sum_\alpha\int \rd\v
\frac{n_\alpha m_\alpha \cv^2}{2} f_\alpha(X,t)\right\} \nonumber\\
&&+ \frac{\partial}{\partial \r}\left\{ \frac{c}{4\piup} \left[\E(\r,t)\times\B(\r,t)\right]+
\sum_\alpha\int \rd\v\, \v
\frac{n_\alpha m_\alpha \cv^2}{2} f_\alpha(X,t)\right\}
+ \J^\text{e}(\r,t)\E(\r,t)=0,
\label{eq2-14}
\end{eqnarray}
where the terms responsible for the particle energy and energy flux are present in the explicit form. We see that there is no need to extract the particle energy term from the quantity $\E\frac{\partial\D}{\partial t}$ as it is done for the derivation of  equation (\ref{eq2-8}) and its generalization to the case of a weakly absorptive medium \cite{10,16,19}.

Thus, the problem under consideration can be solved, if the distribution function is known. On the other hand, equation (\ref{eq2-14}) opens up the possibility to use physical arguments to describe the particle energy  contribution to the energy of perturbation without restriction to the treatment of the case of a transparent medium.

Notice that equation (\ref{eq2-7}) can be easily generalized to the case of a molecular medium, or a combined plasma-molecular medium. To do this in the case of classical model of molecular system, it is necessary to use the kinetic equation for bound pairs \cite{22,23,24}, multiply it by the bound particle energy in terms of the variables describing the internal motion and velocity of center of mass of the bound pair and integrate over these variables.  As a result, the appropriate additional terms taking account of the kinetic and potential energy of bound particles are generated in the equation of the type of equation (\ref{eq2-14}).

\section{Energy density of the electromagnetic field perturbation with regard to the particle energy acquired under the action of the field}  \label{sec3}

In the zero-order approximation on the gas-dynamic parameter ($l/L\ll1$, where $l$ is the mean free path, $L$ is the size of the system) the solution of the kinetic equation (\ref{eq2-4}) may be written in the form of the local Maxwellian distribution \cite{22}
\begin{equation}
f_\alpha(X,t)=\frac{n_\alpha(\r,t)}{n_\alpha} \left[\frac{m_\alpha}{2\piup T_\alpha(\r,t)}\right]^{3/2}\ \exp\left\{{-\frac{m_\alpha[\v-\u_\alpha(\r,t)]^{2}}{2T_\alpha(\r,t)}}\right\},
\label{eq3-1}
\end{equation}
where
\begin{eqnarray}
n_\alpha(\r,t) &=& n_\alpha \int \rd\v f_\alpha(X,t),  \nonumber\\
u_\alpha(\r,t) &=& \frac{n_\alpha \int \rd\v \v f_\alpha(X,t)}{n_\alpha(\r,t)}\,, \nonumber\\
T_\alpha(\r,t) &=& \frac{n_\alpha \int \rd\v (m_\alpha/2)[\v-\u_\alpha(\r,t)]^2f_\alpha(X,t)}{3 n_\alpha(\r,t)}.
\label{eq3-2}
\end{eqnarray}

Within such an approximation, we can present the full energy density as given by
\begin{eqnarray}
W &=& W_\text{F}+W_\text{T}+W_\text{K} \nonumber\\ &=& \frac{1}{8\piup} \Bigl[ \E^2(\r,t)+\B^2(\r,t)\Bigr]
+
\sum_\alpha \left[ \frac{3}{2} n_\alpha(\r,t) T_\alpha(\r,t) + \frac{m_\alpha n_\alpha(\r,t) \u_\alpha^2(\r,t)}{2} \right],
\label{eq3-3}
\end{eqnarray}
where the field $W_\text{F}$, thermal $W_\text{T}$ and kinetic $W_\text{K}$  energies, respectively, are given by
\begin{eqnarray}
W_\text{F} &=& \frac{1}{8\piup} \left[ \E^2(\r,t)+\B^2(\r,t)\right], \nonumber\\
W_\text{T} &=& \sum_\alpha \frac{3}{2} n_\alpha(\r,t) T_\alpha(\r,t), \nonumber\\
W_\text{K} &=& \sum_\alpha n_\alpha(\r,t) \frac{m_\alpha \boldsymbol{u}_\alpha^2(\r,t)}{2}.
\label{eq3-4}
\end{eqnarray}
Since $W_\text{T}$ is the heat produced by the perturbation we can treat the energy associated with the electromagnetic field as the sum of $W_\text{F}$ and $W_\text{K}$.

Restricting ourselves by the second order approximation in the perturbation, we can rewrite the part of energy $W_\text{K}$ as
\begin{equation}
W_\text{K} =\sum_\alpha \frac{n_\alpha m_\alpha \u_\alpha^2(\r,t)}{2} = \sum_\alpha \frac{m_\alpha}{2e_\alpha^2 n_\alpha}\, J_\alpha^2(\r,t).
\label{eq3-5}
\end{equation}
Here, $J_\alpha(\r,t)$ is the partial contribution of the particle of $\alpha$ species to the induced current $J(\r,t)=\sum_\alpha J_\alpha(\r,t)$.

It should be noted that equation (\ref{eq3-5}) directly follows from the transparent physical reasoning: the kinetic energy acquired by particles under the action of the electromagnetic field can be directly expressed in terms of the averaged induced velocity. Namely, this approach was used to estimate the energy density of particles in the case of cold plasmas \cite{2,10,14}. However, as is seen, the equation (\ref{eq3-5}) does not require such restrictions.

Generalization of the results obtained in \cite{2,10,14} can be achieved using the relation between the induced current and the electric field (\ref{eq2-2}) that yields
\begin{eqnarray}
W_\text{K} &= &\sum_\alpha \frac{m_\alpha}{2e_\alpha^2 n_\alpha}\int\limits_{-\infty}^t \rd t' \int \rd\r'  \sigma_{ij}^{(\alpha)}(\r,\r';t-t')
\int\limits_{-\infty}^t \rd t'' \int \rd\r'' \sigma_{ij}^{(\alpha)}(\r,\r'';t,t'') E_j(\r',t')E_\k(\r'',t'') \nonumber\\
&=&\sum_\alpha \frac{m_\alpha}{2e_\alpha^2 n_\alpha} \int \frac{\rd\omega}{2\piup}  \int \frac{\rd\k}{(2\piup)^3} \int \frac{\rd\omega'}{2\piup} \int \frac{\rd\k'}{(2\piup)^3}\nonumber\\
&\times&
\re^{-\ri(\omega-\omega')t} \re^{\ri(\k-\k')\r} \sigma_{ij}^{(\alpha)}(\k,\omega) \sigma_{ik}^{(\alpha)*}(\k',\omega') E_{i\k\omega}E_{j\k'\omega'}^*\, ,
\label{eq3-6}
\end{eqnarray}
where $\sigma_{ij}^{(\alpha)}(\k,\omega)$ is the partial contribution of particles of $\alpha$ species to the conductivity tensor of the system
\[
\sigma_{ij}(\k,\omega) = \sum_\alpha \sigma_{ij}^{(\alpha)}(\k,\omega),
\]
or in terms of the generalized polarizability $\chi_{ij}^{(\alpha)}(\k,\omega)\equiv\frac{4\piup \ri}{\omega} \sigma_{ij}^{(\alpha)}(\k, \omega)$,
the effective energy of electromagnetic perturbation in the medium $W_\text{F}^{\eff}\equiv W_\text{F}+W_\text{K}$
can be written as
\begin{eqnarray}
W_\text{F}^{\eff} &=& \frac{1}{8\piup} \int\frac{\rd\k}{(2\piup)^3}\int\frac{\rd\k'}{(2\piup)^3} \int\frac{\rd\omega}{2\piup}\int\frac{\rd\omega'}{2\piup} \re^{\ri(\k-\k')\r} \re^{-\ri(\omega-\omega')t} \nonumber\\
&\times&\!\!\!\!\Biggl[ \frac{k_ik'_j}{\k\k'} +\left(1+\frac{c^2}{\omega\omega'} \k\k'\right) \left(\delta_{ij}-\frac{k_i k'_j}{\k\k'}\right) +\sum_{\alpha=\text{e,i}} \frac{\omega^2}{\omega_{\text{p}\alpha}^2} \chi_{ki}^{(\alpha)}(\k,\omega)\chi_{k j}^{(\alpha)*}(\k',\omega')\Biggr] E_{i\k\omega} E_{j\k'\omega'}^*\, ,
\label{eq3-7}
\end{eqnarray}
where $\omega_{\text{p}\alpha}^2=4\piup e_\alpha^2 n_\alpha/m_\alpha$.

This is the general relation for the electromagnetic perturbation energy in a plasma-like medium.

It should be noted that equation (\ref{eq3-5}) can be also used to estimate the kinetic energy of bound electrons in atoms and molecules. However, in this case the energy of electromagnetic perturbation  along with the kinetic energy of electrons also contains the potential energy of bound electrons in the fields of ions with which they are bound. In the case of the classical model of the atom-oscillator \cite{10,22,27} such energy can be estimated as
\[
W_\text{U} = n_\text{m} \frac{\omega_0^2 r_\text{m}^2(\r,t)}{2}.
\]
Here, $n_\text{m}$ is the density of the bound electrons, $\omega_0$ is the eigenfrequency of the oscillator, $\r_\text{m}(\r,t)$ is the reduced coordinate of the bound electron. Since $\u_\text{m}(\r,t)=\frac{\rd\r_\text{m}(\r,t)}{\rd t}$, the energy $W_\text{U}$ may be expressed in terms of the mean velocity $\u_\text{m}(\r,t)$, i.e., in terms of the induced current of the bound electrons.
\textit{The validity of the expressions for kinetic and potential energy of the bound particles   is confirmed by the appropriate derivation on the basis of the equation of the type of equations} (\ref{eq2-14})--(\ref{eq3-2}) \textit{for a classical molecular model.} Thus,
\begin{eqnarray}
W_\text{U} &=& \frac{1}{8\piup} \int \frac{\rd\k}{(2\piup)^3} \int\frac{\rd\k'}{(2\piup)^3} \int\frac{\rd\omega}{2\piup} \int\frac{\rd\omega'}{2\piup} \, \re^{\ri(\k-\k')\r}\nonumber\\
&\times&  \re^{-\ri(\omega-\omega')t} \frac{\omega_0^2}{\omega_\text{pm}^2} \chi_{ki}^{(\text{m})}(\k,\omega) \chi_{lj}^{(\text{m})*}(\k',\omega') E_{i\k\omega} E_{j\k'\omega'}^*\,,
\label{eq3-8}
\end{eqnarray}
where $\chi_{ij}^{(\text{m})}(\k,\omega)$ in the case of the classical model of an atom-oscillator is given by \cite{23}
\begin{equation}
\chi_{ij}^{(\text{m})}(\k,\omega) =  -\, \delta_{ij}\int \rd\v \frac{\omega_\text{pm}^2 f_{0\text{m}}(\v)}{(\omega-\k\v)^2-\omega_0^2+\ri\gamma(\omega-\k\v)}\,,
\label{eq3-9}
\end{equation}
$\omega_\text{pm}^2 = 4\piup e_\text{b}^2 n_\text{m}/m_\text{b}$, $f_{0\text{m}}(\v)$ is the distribution function of bound particles (atoms, or molecules), $e_\text{b}$ and $m_\text{b}$ are the effective charge and the reduced mass of a bound electron.

Thus, in the case of a plasma-molecular system, the energy of a perturbation may be written as
\begin{eqnarray}
W &=&  W_\text{F}+W_\text{K}+W_\text{U}= \frac{1}{8\piup} \int\frac{\rd\k}{(2\piup)^3} \int\frac{\rd\k'}{(2\piup)^3} \int\frac{\rd\omega}{2\piup} \int\frac{\rd\omega'}{2\piup} \, \re^{\ri(\k-\k')\r} \;\re^{-\ri(\omega-\omega')t}  \nonumber\\
&\times& \Biggl[ \left(\delta_{ij}-\frac{k_i k'_j}{\k\k'}\right) \left(1+\frac{c^2\k\k'}{\omega^2}\right) +\frac{k_i k'_j}{\k\k'} +
 \sum_{\alpha=\text{e,i}} \frac{\omega^2}{\omega_{\text{p}\alpha}^2} \chi_{ki}^{(\alpha)}(\k,\omega) \chi_{kj}^{(\alpha)*}(\k',\omega')\nonumber\\
&+ & \frac{\omega^2+\omega_0^2}{\omega_\text{pm}^2} \, \chi_{ki}^{(\text{m})}(\k,\omega) \chi_{kj}^{(\text{m})*}(\k',\omega') \Biggr]
 E_{i\k\omega} E_{j\k'\omega}^*\,.
\label{eq3-10}
\end{eqnarray}

This equation remains valid in the case of quantum description provided the polarizabilities  $\chi_{ij}^{(\alpha)}(\k,\omega)$ ($\alpha=\text{e,\,i,\,m}$) are calculated appropriately (see, for example, \cite{22,24}).

In the case of the monochromatic field $\E(\r,t)={1}/{2} \left[ \E(\r)\re^{-\ri\omega t} +\E^*(\r) \re^{\ri\omega t}\right]$, after the averaging over the oscillation period $T=2\piup/\omega$ and the volume of the system $V$, equation (\ref{eq3-10}) is reduced to
\begin{eqnarray}
\overline{W} &=& \frac{1}{16\piup V} \int\frac{\rd\k}{(2\piup)^3} \Biggl[ \left(\delta_{ij}-\frac{k_i k_j}{k^2}\right)\left(1+\frac{c^2k^2}{\omega^2}\right) +\frac{k_ik_j}{k^2} \nonumber\\
&+&\sum_{\alpha=\text{e,i}} \frac{\omega^2}{\omega_{\text{p}\alpha}^2} \, \chi_{ki}^{(\alpha)}(\k,\omega) \chi_{kj}^{(\alpha)*}(\k,\omega)
+ \frac{\omega^2+\omega_0^2}{\omega_\text{pm}^2}\, \chi_{ki}^{(\text{m})}(\k,\omega) \chi_{kj}^{(\text{m})*}(\k,\omega)\Biggr] \ \overline{E_{\k i}E_{\k j}^*}.
\label{eq3-11}
\end{eqnarray}

Let us consider the case when the spatial dispersion is formally neglected, i.e., we put
$\chi_{ij}^{(\alpha)}(\k,\omega) = \chi_{ij}^{(\alpha)}(\omega)$. In this case, equation (\ref{eq3-11}) is simplified to
\begin{equation}
\overline{W}\equiv\overline{W}_E+\overline{W}_B\,,
\label{eq3-12}
\end{equation}
where
\begin{eqnarray}
\overline{W}_E &=& \frac{1}{16\piup} \Biggl[ \delta_{ij}+\sum_{\alpha=\text{e,i,m}} \frac{\omega^2+\omega_{0\alpha}^2}{\omega_{\text{p}\alpha}^2}\,
\chi_{ki}^{(\alpha)}(\omega) \chi_{kj}^{(\alpha)*}(\omega)\Biggr]
\ \overline{E_i E^*_j}\,,\nonumber\\
\overline{W}_B &=& \frac{1}{16\piup}\overline{|\B|^2} ,\quad \overline{|\B|^2} =\frac{1}{V} \int \rd\r |\B(\r)|^2, \nonumber\\
\overline{E_i E^*_j} &=& \frac{1}{V} \int \rd\r E_i(\r) E_j^*(\r), \nonumber\\
\omega_{0\alpha} &=&
\left\{ \begin{array}{cc}
0, & \alpha=\text{e,\,i},\\
\omega_0, & \alpha=\text{m}. \end{array} \right.\, 
\label{eq3-13}
\end{eqnarray}

Using (\ref{eq3-12}), (\ref{eq3-13}) it is easy to recover the results obtained in \cite{10,11,12} for the electric field energy density outside the transparency domain. For example, in the case of a cold molecular system
\begin{equation}
\chi_{ij}^{(\text{m})}(\omega) =-\delta_{ij}\frac{\omega_\text{pm}^2}{\omega^2-\omega_0^2+\ri\gamma\omega}\,,
\label{eq3-14}
\end{equation}
which leads to
\begin{equation}
\overline{W}_E =\frac{1}{16\piup} \overline{|\E|^2} \left[ 1+\frac{\omega_\text{pm}^2 (\omega^2+\omega_0^2)}{(\omega^2-\omega_0^2)^2+\gamma^2\omega^2}\right].
\label{eq3-15}
\end{equation}

In the case of a cold plasma
\begin{equation}
\chi_{ij}^{(\text{e})}(\omega) =-\delta_{ij}\frac{\omega_\text{pe}^2}{\omega(\omega+\ri\nu_\text{e})}\,,
\label{eq3-16}
\end{equation}
where $\nu_\text{e}$ is the effective collision frequency, that gives
\begin{equation}
\overline{W}_E=\frac{1}{16\piup} \left( 1+\frac{\omega_\text{pe}^2}{\omega^2+\nu_\text{e}}\right) \overline{|\E|^2}.
\label{eq3-17}
\end{equation}
Equations (\ref{eq3-15}) and (\ref{eq3-17}) are in agreement with the well-known relation [see the first term in  equation~(\ref{eq2-10})]
\begin{equation}
\overline{W} =\frac{1}{16\piup} \frac{\partial}{\partial\omega} \left[\omega \Re\varepsilon_{ij}(\omega)\right] \overline{E_i E_j^*}
\label{eq3-18}
\end{equation}
only in the case of nondissipative systems ($\gamma=0$ and $\nu_\text{e}=0$).

\section{Energy density of the electromagnetic field fluctuations}   \label{sec4}

Within the context of the theory of electromagnetic fluctuations it is easy to show that equation (\ref{eq3-10})  may be also applied to the description of the energy density  of fluctuations. The statistical averaging of equation (\ref{eq3-10}) yields
\begin{eqnarray}
\langle W\rangle &=& \frac{1}{8\piup} \int\frac{\rd\k}{(2\piup)^3} \int\frac{\rd\omega}{2\piup} \Biggl[ \frac{k_i k_j}{k^2} +
\left(1+\frac{c^2k^2}{\omega^2}\right)
\left( \delta_{ij}-\frac{k_i k_j}{k^2}\right) \nonumber\\
&+& \sum_{\alpha=\text{e,i,m}} \frac{\omega^2+\omega_{0\alpha}^2}{\omega_{\text{p}\alpha}^2}\, \chi_{ki}^{(\alpha)}(\k,\omega) \chi_{kj}^{(\alpha)*}(\k,\omega) \Biggr] \langle \delta E_i \delta E_j\rangle_{\k\omega}.
\label{eq4-1}
\end{eqnarray}
When deriving equation (\ref{eq4-1}) we take into account that
\begin{equation}
\langle\delta E_{i\k\omega} \delta E_{j\k'\omega'}^*\rangle = (2\piup)^4 \delta(\k-\k') \delta(\omega-\omega') \langle \delta E_i \delta E_j\rangle_{\k\omega}\,,
\label{eq4-2}
\end{equation}
where
\begin{eqnarray}
\langle \delta E_i \delta E_j\rangle_{\k\omega} &=& \int \! \rd\R \,\re^{-\ri\k\R}\! \int \rd\omega\, \e^{\ri\omega\tau} \langle \delta E_i(\r,t) \delta E_j(\r',t')\rangle_{\k\omega}\,, \nonumber\\
&&\R =  \r-\r', \qquad \tau=t-t'.
\label{eq4-3}
\end{eqnarray}

In the case of an equilibrium system $\langle \delta E_i \delta E_j\rangle_{\k\omega}$ is given by the fluctuation dissipation theorem (see, for example, \cite{3,4})
\begin{equation}
\langle \delta E_i \delta E_j\rangle_{\k\omega} =\frac{4\piup \ri}{\omega} \theta(\omega) \left[\Lambda_{ij}^{-1}(\k,\omega)-\Lambda_{ji}^{-1*}(\k,\omega)\right].
\label{eq4-4}
\end{equation}
Here,
\begin{eqnarray}
\theta \equiv   \frac{\hbar\omega}{2}\coth \frac{\hbar\omega}{2T}\,, \qquad
\Lambda_{ij}(\k,\omega) = \varepsilon_{ij}(\k,\omega) -\frac{k^2c^2}{\omega^2}\left(\delta_{ij}-\frac{k_ik_j}{k^2}\right).
\label{eq4-5}
\end{eqnarray}

Further simplification of (\ref{eq4-1}), (\ref{eq4-5}) can be done in the case of an isotropic system for which
\begin{equation}
\varepsilon_{ij}(\k,\omega) =\varepsilon_\T(k,\omega) \left(\delta_{ij}-\frac{k_ik_j}{k^2}\right) +\varepsilon_\L(k,\omega) \frac{k_ik_j}{k^2}\,,
\label{eq4-6}
\end{equation}
where $\varepsilon_\T(k,\omega)$ and $\varepsilon_\L(k,\omega)$ are the transverse and longitudinal parts of the dielectric permittivity tensor.

Substituting (\ref{eq4-6}) into (\ref{eq4-4}) and (\ref{eq4-1}) yields
\begin{equation}
\langle W\rangle =\int\limits_0^\infty \langle W\rangle_\omega \rd\omega,
\label{eq4-7}
\end{equation}
where for the general case of the non-transparent medium we have
\begin{eqnarray}
\langle W\rangle_\omega &=& \frac{\theta(\omega)}{2\piup^3\omega}\int\limits_0^\infty \rd k\, k^2 \Biggl\{ \frac{\Im\varepsilon_\L(k,\omega)}{|\varepsilon_\L(k,\omega)|^2}
 \Biggl[1+ \sum_{\alpha=\text{e,i,m}}\frac{\omega^2+\omega_{0\alpha}^2}{\omega_{\text{p}\alpha}^2} |\chi_\L^{(\alpha)}(k,\omega)|^2\Biggr] \nonumber\\
&+& \frac{2\Im\varepsilon_\T(k,\omega)}{|\varepsilon_\T(k,\omega)-k^2c^2/\omega^2|^2}
 \Biggl[ 1+ \frac{k^2c^2}{\omega^2} +\sum_{\alpha=\text{e,i,m}} \frac{\omega^2+\omega_{0\alpha}^2}{\omega_{\text{p}\alpha}^2} |\chi_\T^{(\alpha)}(k,\omega)|^2\Biggr]\Biggr\}\, ,
\label{eq4-8}
\end{eqnarray}
that describes the contribution of both longitudinal and transverse electromagnetic fields.

In the case of negligible dissipation, we can use the approximation of the type
\begin{eqnarray}
\frac{\Im \varepsilon_\T(k,\omega)}{|\varepsilon_\T(k,\omega)-k^2c^2/\omega^2|^2}  \simeq  \piup\delta \left(\Re\varepsilon_\T(k,\omega)-\frac{k^2c^2}{\omega^2}\right).
\label{eq4-9}
\end{eqnarray}

In the case of cold plasma for $\omega\gg\nu$ we have
\[
\varepsilon_\T(\omega)\simeq 1-\frac{\omega_\text{p}^2}{\omega^2}
\]
and the part of the energy associated with the transverse electromagnetic field is given by
\begin{eqnarray}
\langle W\rangle_\omega =\frac{\omega^2\theta(\omega)}{\piup^2 c^3}\sqrt{\varepsilon(\omega)}, \qquad (\omega>\omega_\text{p}).
\label{eq4-10}
\end{eqnarray}
This relation is in agreement with the well-known result for the energy density in the dispersive transparent medium \cite{29} and reproduces the energy density for transparent plasmas \cite{30}.

In the case of a molecular medium with weak absorption $\gamma\to0$
\begin{equation}
\varepsilon_\text{b}(\omega) =1-\frac{\omega_\text{pm}^2}{\omega^2-\omega_0^2}
\label{eq4-11}
\end{equation}
and thus,
\[
\langle W\rangle_\omega = \frac{\omega^2\theta(\omega)}{2\piup^2 c^3}\sqrt{\Re\varepsilon_\text{b}(\omega)} \left[1+\Re\varepsilon_\text{b}(\omega)+\frac{\omega_\text{pm}^2(\omega^2+\omega_0^2)}{(\omega^2-\omega_0^2)^2+\gamma^2\omega^2}\right]\, .
\]
For $\omega\gg\omega_0$ we come back to the equation of the type (\ref{eq4-10}).

For $\omega\ll\omega_0$, the frequency dispersion can be neglected and we obtain the result for nondispersive transparent medium \cite{29}
\begin{equation}
\langle W\rangle_\omega = \frac{\omega^2\theta(\omega)}{2\piup^2 c^3}\tilde{\varepsilon}^{3/2} ,
\label{eq4-12}
\end{equation}
where
\[
\tilde{\varepsilon}=\lim\limits_{\omega\to 0}\, \varepsilon_\text{b}(\omega).
\]

In the general case, equation (\ref{eq4-8}) may be rewritten in the form
\begin{equation}
\langle W\rangle_\omega = \frac{\hbar\omega^3}{\piup^2 c^3}\left( \frac{1}{2}+\frac{1}{\e^{\hbar\omega/T}-1}\right) S(\omega),
\label{eq4-13}
\end{equation}
where $S(\omega)$ is the function describing the effect of the medium
\begin{eqnarray}
S(\omega) &=& \frac{c^3}{2\piup\omega^3} \int\limits_0^\infty \rd k\, k^2 \Biggl\{ \frac{\Im \varepsilon_\L(k,\omega)}{|\varepsilon_\L(k,\omega)|^2}
 \left[ 1+\sum_{\alpha=\text{e,i,m}}\frac{\omega^2+\omega^2_{0\alpha}}{\omega^2_{\text{p}\alpha}} |\chi_\L^{(\alpha)}(k,\omega)|^2\right]\nonumber\\
&+& \frac{2\Im \varepsilon_\T(k,\omega)}{|\varepsilon_\T(k,\omega)-k^2c^2/\omega^2|^2}
 \Biggl[ 1+\frac{k^2c^2}{\omega^2}
+ \sum_{\alpha=\text{e,i,m}}\frac{\omega^2+\omega^2_{0\alpha}}{\omega^2_{\text{p}\alpha}} |\chi_\T^{(\alpha)}(k,\omega)|^2\Biggr]\Biggr\}.
\label{eq4-14}
\end{eqnarray}

\section{Numerical analysis of the fluctuation field energy spectrum}   \label{sec5}

As mentioned above, equations (\ref{eq4-8}) and (\ref{eq4-14}) take into account the contribution of both longitudinal and transverse electromagnetic field fluctuations. Particularly, the contribution of transverse fields into the function $S(\omega)$ is given by
\begin{equation}
S_\text{T}(\omega) =\frac{c^3}{\piup\omega^3}\int\limits_0^{\infty} \rd k\, k^2
\frac{\Im \varepsilon_\text{T}(k,\omega)}{| \varepsilon_\text{T}(k,\omega ) -
k^2 c^2/\omega^2|^2} \left[1+\frac{k^2 c^2}{\omega^2}+\sum_{\alpha=\text{e,i,m}}
\frac{\omega^2+\omega_0^2}{\omega_{\text{p}\alpha}^2}| \chi_\text{T}^{(\alpha)}(k,\omega) |^2 \right],
\label{eq5-1}
\end{equation}
and thus,
\begin{equation}
\langle W\rangle_\omega^\T= \frac{\hbar\omega^3}{\piup^2c^3} \left(\frac{1}{2}+\frac{1}{\re^{\hbar\omega/T}-1}\right) \, S_\text{T}(\omega)
\label{eq5-2}
\end{equation}
can be considered as a generalization of the Planck formula to the case of absorptive medium. The function $S_\text{T}(\omega)$ makes it possible to see what is the influence of the medium on the equilibrium radiation.

Particularly, in the case of a transparent medium without spatial dispersion [$\varepsilon_\text{T}(k,\omega)\equiv\varepsilon_\text{T}(\omega)$]
\[
S_\text{T}(\omega)=\sqrt{\varepsilon(\omega)}
\]
and in the case of transparent nondispersive medium [$\varepsilon_\text{T}(\omega)\equiv\varepsilon_\text{T}$]
\[
S_\text{T}(\omega)=\varepsilon_\text{T}^{3/2}.
\]

Now, let us consider the results of numerical calculation of the quantity $S_\text{T}(\omega)$ in the case of collisional plasmas.

The calculations have been performed for one-component classical electron plasma, i.e.,
\begin{eqnarray}
S_\text{T}(\omega) =\frac{c^3}{\piup\omega^3}\int\limits_0^{\infty} \rd k\, k^2
\frac{\Im \varepsilon_\text{T}(k,\omega)}{| \varepsilon_\text{T}(k,\omega ) -
k^2 c^2/\omega^2|^2}
 \left[ 1+\frac{\omega^2}{\omega_\text{p}^2}\left|\chi_\text{T}(k,\omega)\right|^2 +\frac{k^2c^2}{\omega^2}\right].
\label{eq5-3}
\end{eqnarray}
The quantity $\chi_\text{T}(k,\omega)$ was calculated on the basis of the Bhatnagar-Gross-Krook model
\[
 \chi_\text{T}(k,\omega )=-\frac{\omega_\text{p}^2}{\omega(\omega +\ri\nu)}\left[1-W \left(\frac{\omega +\ri\nu}{kv_\text{T}} \right)\right],
 \]
where $v_\text{T}^2=T_\text{e}/m_\text{e}$, $\nu$ is the collision frequency, $W(z)$ is the plasma dispersion function
\[
W(z)=1-z\e^{-z^2/2} \int\limits_0^z\, \rd y\,\e^{y^2/2} +\ri\left(\frac{\piup}{2}\right)^{1/2} z\e^{-z^2/2}\, .
\]
$S_\text{T}(\omega)$ can be divided into two parts
\begin{equation}
S_\text{T}^{\EK}(\omega)=\frac{c^3}{\piup\omega^3}\int\limits_0^\infty \rd k\, k^2 \frac{\Im \varepsilon_\text{T}(k,\omega)}{|\varepsilon_\text{T}(k,\omega) -k^2 c^2/\omega^2|^2} \left[1+\frac{\omega^2}{\omega_\text{p}^2}\left|\chi_\text{T}(k,\omega)\right|^2 \right]
\label{eq5-4}
\end{equation}
and
\begin{equation}
S_\text{T}^{B}(\omega)=\frac{c^5}{\piup\omega^5}\int\limits_0^\infty \rd k\, k^4 \frac{\Im \varepsilon_\text{T}(k,\omega)}{|\varepsilon_\text{T}(k,\omega) -k^2 c^2/\omega^2|^2}.
\label{eq5-5}
\end{equation}
The first one is associated with the electric field energy and particle kinetic energy, and the second one with the magnetic field fluctuations.

$S_\text{T}^{B}(\omega)$ in its turn can be presented as
\begin{equation}
S_\text{T}^B(\omega)=S_\text{T}^{B\rdiv}(\omega)+S_\text{T}^{B\,\reg}(\omega),
\label{eq5-6}
\end{equation}
where
\begin{equation}
S_\text{T}^{B\rdiv} =\frac{c}{\piup\omega}\int\limits_0^\infty \rd k\, \Im \varepsilon_\text{T}(k,\omega),
\label{eq5-7}
\end{equation}
\begin{equation}
S_\text{T}^{B\,\reg} =\frac{c}{\piup\omega}\int\limits_0^\infty \rd k\, \Im \left[-\frac{\varepsilon_\text{T}^2(k,\omega)}{\varepsilon_\text{T}-k^2c^2/\omega^2}\right].
\label{eq5-8}
\end{equation}

As it easy to show, the integrals in equations (\ref{eq5-4}) and (\ref{eq5-8}) are convergent. On the contrary, the integral in equation (\ref{eq5-7}) with $\varepsilon_\text{T}(k,\omega)$ under consideration is divergent and the restriction of integration to the cut-off wavenumber $k_{\max}$ is needed. This problem is well-known and has been discussed in detail in \cite{31,32,33}. Notice that in the case of quantum calculations of $\varepsilon_\text{T}(k,\omega)$, the integral in equation (\ref{eq5-7}) is convergent which follows from the explicit form of $\Im \varepsilon_\text{T}(k,\omega)$ in such a case \cite{34}. However, in the case of classical calculations, the problem of the choice of $k_{\max}$ arises. In order to obtain the qualitative picture of the energy spectrum behaviour, we take the simplest well-known approximation $k_{\max}=T/e^2$.

Herein below we present the results of calculations of $S_\text{T}(\omega)$, $S_\text{T}^{\EK}(\omega)$, $S_\text{T}^{B\rdiv}$, and $S_\text{T}^{B\,\reg}$ for three kinds of plasmas: solar corona, high-pressure gas discharge and electron-positron plasma at the beginning of primordial nucleosynthesis. The parameters of plasmas are presented in table~\ref{tab1}. The previous expressions are reduced to a dimensionless form. In this manner, the energy density depends on dimensionless parameters $w/w_\text{p}$, $c/v_\text{T}$, $\nu/\omega_\text{p}$. The dimensionless value of $k_{\max}$ is
\begin{equation}
 y_\max=\frac{c T}{\omega_\text{pe}e^2}.
\label{eq5-9}
\end{equation}

\begin{table}[!b]
\centering
\caption{Plasma parameters.}
\vspace{2ex}
\begin{tabular}{|l|c|c|c|c|}
  \hline\hline
                           & $n_\text{e}$, cm$^{-3}$ & $T_\text{e}$, eV & $y_\text{max}$ & $c/v_\text{T}$ \\ \hline\hline
  solar corona             & $10^{10}$        & 150       & $5\cdot10^9$   & 40 \\ \hline
  high-pressure gas discharge  & $10^{16}$        & 5         & $1.8\cdot10^5$ & 225 \\ \hline
  electron-positron plasma &                  &           &                &     \\
  at the beginning of primordial nucleosynthesis & $4.8\cdot10^{30}$ & 8.6$\cdot10^5$& 1450           & 1.08 \\ \hline\hline
\end{tabular}
\label{tab1}
\end{table}

\begin{figure}[!t]
\centerline{\includegraphics[width=0.5\textwidth]{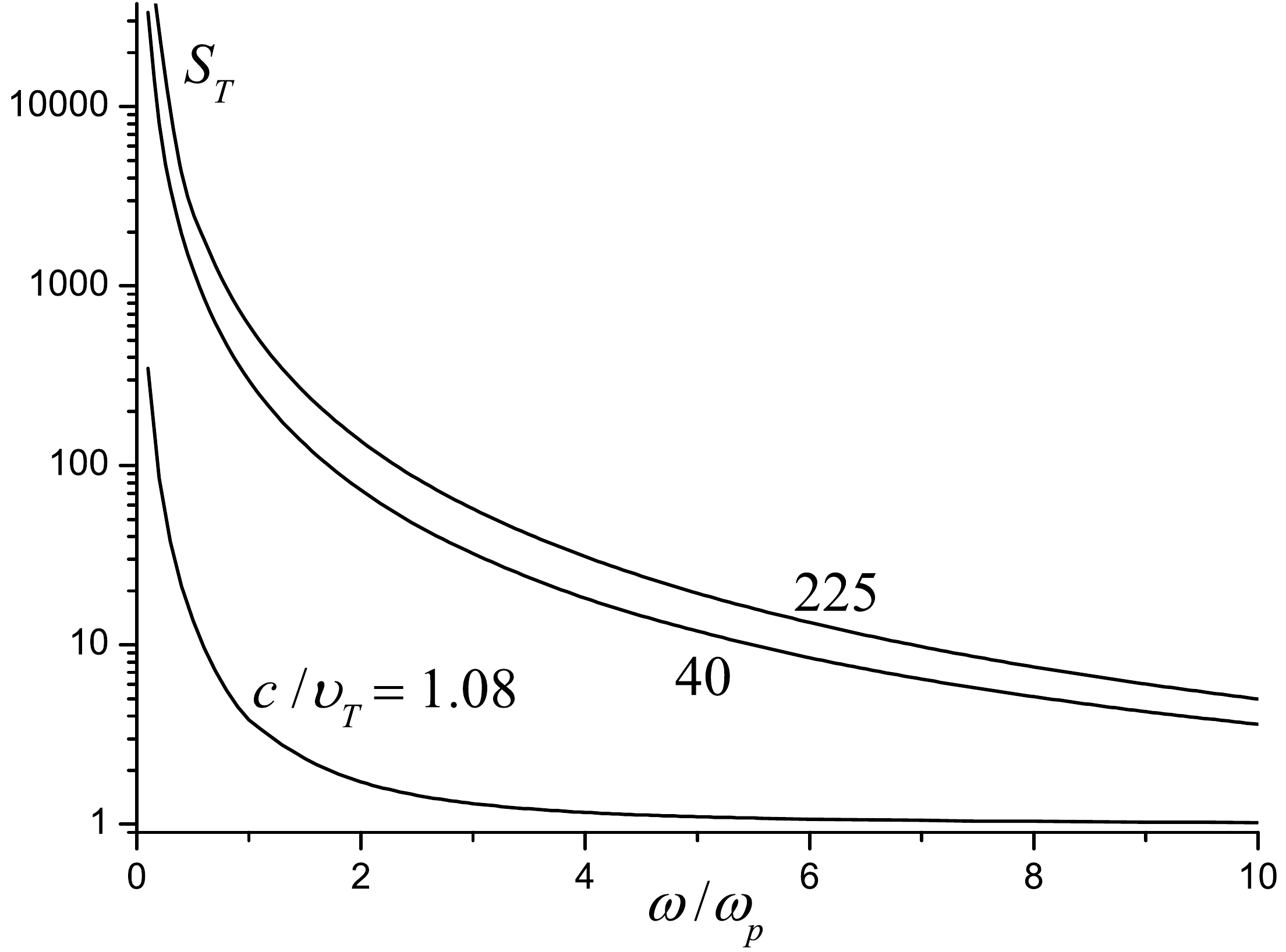}}
\caption{$S_\text{T}$ v.s. $\omega/\omega_\text{p}$ for three types of plasma $c/v_\text{T}=1.08$, 40, 225, $\nu/\omega_\text{p}=0.01$.}\label{fig-logS}
\end{figure}

The total spectral energy density for three types of plasma is shown in figure \ref{fig-logS}. One can see that $S_\text{T}$ considerably depends  on the parameter $c/v_\text{T}$. With an increase of frequency, $S_\text{T}$ tends to a unit which means that for $\omega>\omega_\text{p}$ the energy spectrum reproduces the Planck spectrum.

\begin{figure}[!t]
\centerline{\includegraphics[width=0.5\textwidth]{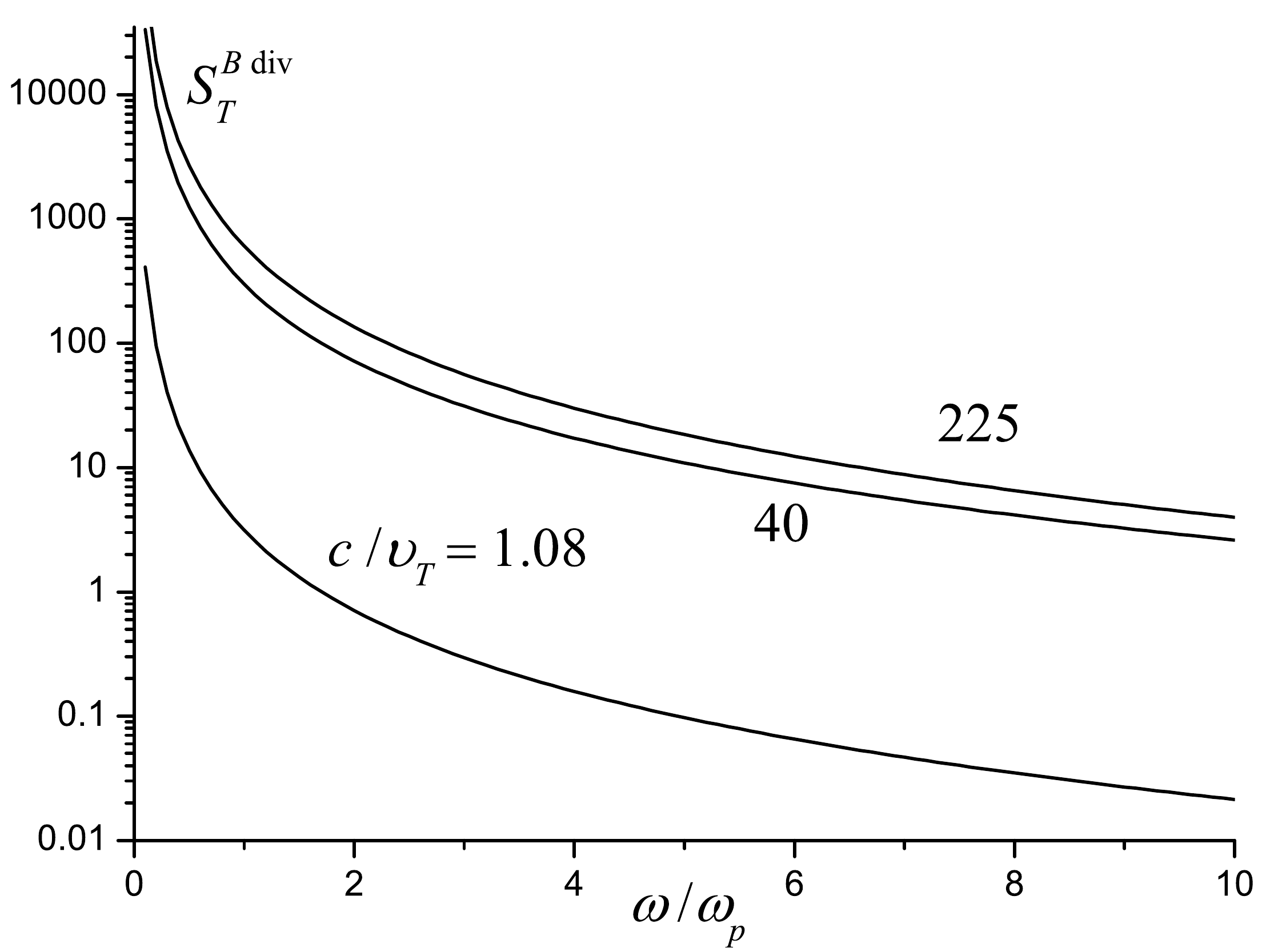}}
\caption{$S_\text{T}^{B\rdiv}$ v.s. $\omega/\omega_\text{p}$ for three types of plasma $c/v_\text{T}=1.08$, $40$, $225$, $\nu/\omega_\text{p}=0.01$.}\label{fig-SBdiv}
\end{figure}

The comparison of figures \ref{fig-logS} and \ref{fig-SBdiv} shows that the energy density is provided almost fully by a divergent part of the magnetic field fluctuations energy density $S_\text{T}^{B\rdiv}$, except the domain of high frequencies, where the divergent part tends to zero.

\begin{figure}[!t]
\centerline{\includegraphics[width=0.45\textwidth]{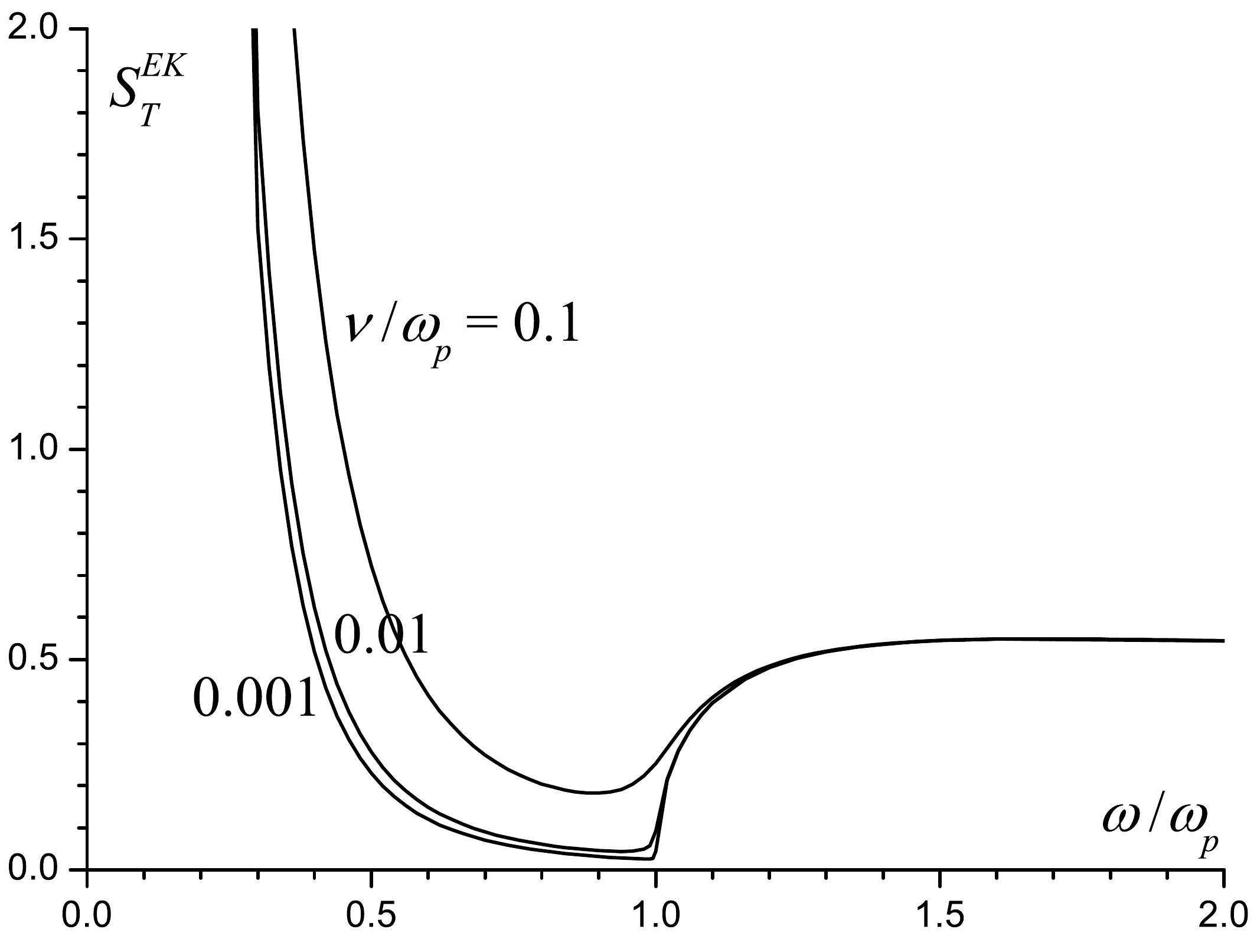}}
\caption{$S_\text{T}^{\EK}$ v.s. $\omega/\omega_\text{p}$ for solar corona $c/v_\text{T}=40$, $\nu/\omega_\text{p}=0.1$, $0.01$, $0.001$.}\label{fig-SEK-korona}
\end{figure}

\begin{figure}[!t]
\centerline{\includegraphics[width=0.45\textwidth]{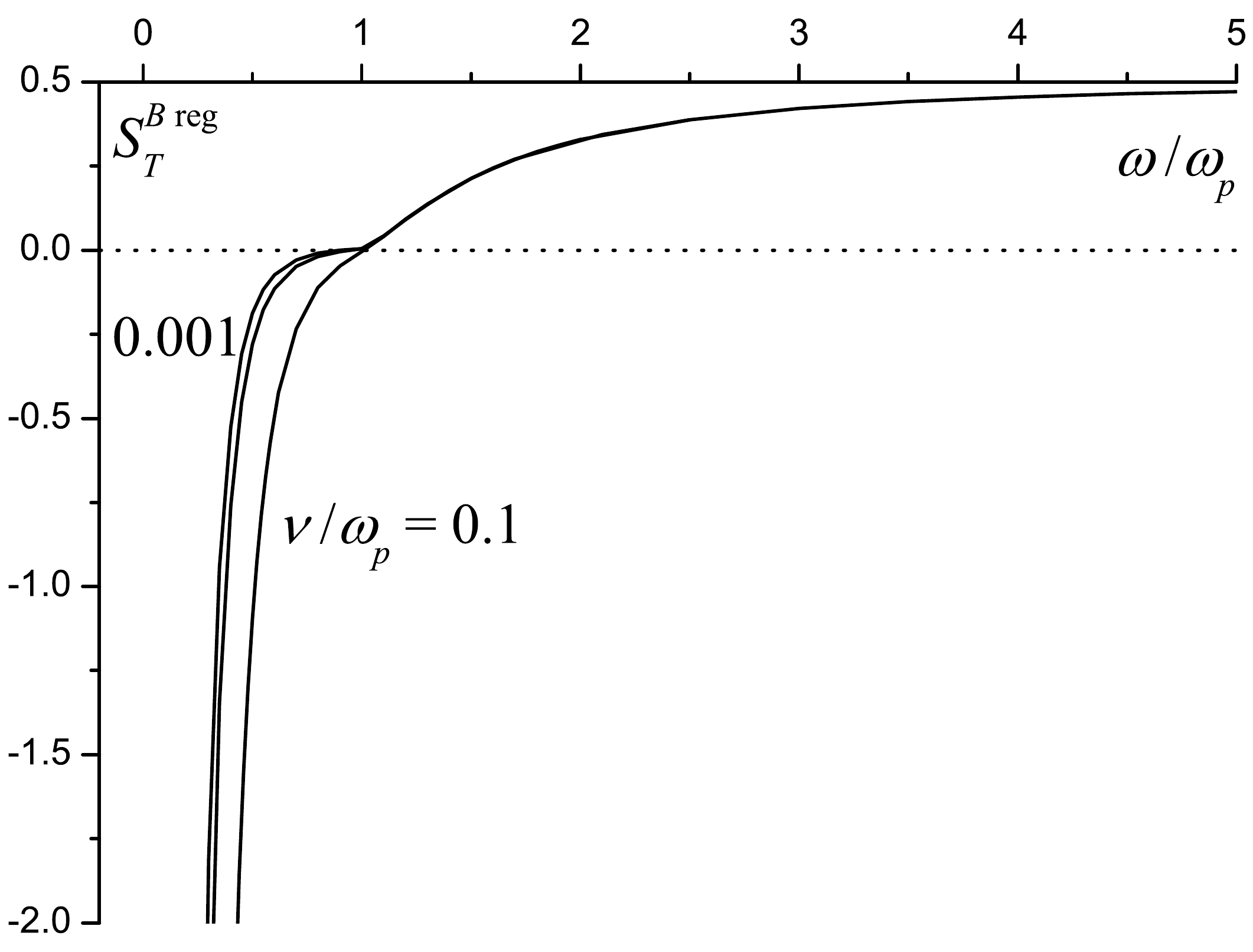}}
\caption{$S_\text{T}^{B\,\reg}$ v.s. $\omega/\omega_\text{p}$ for solar corona $c/v_\text{T}=40$, $\nu/\omega_\text{p}=0.001$, $0.01$, $0.1$.}\label{fig-SBreg-korona}
\end{figure}

$S_\text{T}^{\EK}$ for solar corona is presented in figure \ref{fig-SEK-korona}, and $S_\text{T}^{B\,\reg}$ for solar corona is presented in figure \ref{fig-SBreg-korona}. These quantities depend on the collision frequency, but only for $\omega<\omega_\text{p}$.  Both $S_\text{T}^{\EK}$ and $S_\text{T}^{B\,\reg}$ tend to $0.5$ for high frequency, that corresponds to  the value of total energy density (see figure \ref{fig-logS}). Notice that for the given plasma, parameters $S_\text{T}^{\EK}$ and $S_\text{T}^{B\,\reg}$ with a rather high accuracy do not depend on the cut-off wavenumber.

As it follows from the numerical results, the dissipation of electromagnetic field results in crucial changes of the energy spectrum in the domain in which transverse electromagnetic perturbations cannot propagate (see, figures~\ref{fig-SBdiv}, \ref{fig-SEK-korona}, \ref{fig-SBreg-korona}).  It looks as if in such a case the energy of fluctuating electromagnetic fields can be accumulated in the nontransparent frequency domain, since the generated fields cannot be efficiently emitted from the region of their generation (the radiation mechanism of the energy loss does not work).

This is in a distinct contrast to the case of transparent medium for which
\begin{equation*}
S_\text{T}^\EK (\omega) = \frac{1}{2} \sqrt{\varepsilon_\text{T}(\omega)} \left[ 2-\varepsilon_\text{T}(\omega)\right] \theta(\omega-\omega_\text{p}),
\end{equation*}
\begin{equation*}
S_\text{T}^{B\,\reg}(\omega)=\frac{1}{2} [\varepsilon_\text{T}(\omega)]^{3/2} \theta(\omega-\omega_\text{p}),
\end{equation*}
\begin{equation*}
S_\text{T}^{B\rdiv}(\omega)=0
\end{equation*}
and thus $S_\text{T}(\omega)=\sqrt{\varepsilon_\text{T}(\omega)}\theta(\omega-\omega_\text{p})$.

One more essential feature of the energy spectrum outside  the transparency domain is that the dominant contribution to the spectrum is given by $S_\text{T}^{B\rdiv}(\omega)$, i.e., by the part directly related to the $\Im\varepsilon_\text{T}(\omega)$. Thus, the role of the cut-off wavenumber is very important (as is seen from figures~\ref{fig-logS}, \ref{fig-SBdiv}). Notice that  appropriate calculations show that the magnetic fluctuations still dominate in the case of $\varepsilon_\text{T}(\omega)$ calculated within the quantum approach. It can be explained by the fact that in such a case $\Im\varepsilon_\text{T}(\omega)$ decreases considerably only at $k>\sqrt{2mT/\hbar}$.

\section{Conclusions}

Thus, in the present contribution we derive  general relations for the electromagnetic-field energy density in an absorptive medium with temporal and spatial dispersion. The treatment is based on the assumption that the energy density of an electromagnetic perturbation contains both the electromagnetic field energy and the particle energy acquired in the perturbation field. The results obtained provide a possibility to generalize the Planck law to the case of an absorptive dispersive medium.  The analysis shows that outside  the transparency domain, the dominant contribution to the energy spectrum is given by the magnetic field fluctuations.

The present work was supported by the National Academy of Sciences of
Ukraine within the project ``Mathematical models of nonequilibrium processes
in open systems'' N 0120U100857.

\ukrainianpart

\title{Енергія електромагнетного поля в поглинальному середовищі з часовою і просторовою дисперсією}
\author{А.Г. Загородній\refaddr{label1}, С.А. Трігер\refaddr{label2}, А.І. Момот\refaddr{label3}}
\addresses{
\addr{label1} Інститут теоретичної фізики ім. М.М. Боголюбова Національної академії наук України, \\вул. Метрологічна, 14-б, 03143 Київ, Україна
\addr{label2} Об'єднаний інститут високих температур РАН, вул. Ізгорська, 13, 125412 Москва, Росія
\addr{label3} Київський національний університет імені Тараса Шевченка, \\вул. Володимирська, 64/13, 01601 Київ, Україна
}

\makeukrtitle

\begin{abstract}
\tolerance=3000%
Отримано загальні співвідношення для енергії електромагнетного поля поза областю прозорости. Показано, що внесок заряджених частинок в енергію електромагнетних збурень у загальному випадку можна описати через білінеарну комбінацію діелектричної поляризованости середовища. Знайдено явний вигляд такого внеску. Отримані співвідношення використано для узагальнення закону Планка на випадок поглинального середовища.

\keywords часова і просторова дисперсія, поглинальне середовище, енергія електромагнетного поля

\end{abstract}

\end{document}